\documentclass[letterpaper, oneside]{article}
\usepackage[margin=2.5cm]{geometry}
\usepackage{amsmath}
\usepackage{epsfig}
\usepackage{tensor}
\usepackage{amsfonts}
\usepackage{subfig}
\usepackage{setspace}
\usepackage{hyperref}
\usepackage{authblk}

\newcommand{\ket}[1]{\ensuremath{\mid#1 \rangle}}

\newcommand{\braket}[2]{\ensuremath{\langle#1\mid#2 \rangle}}

\newcommand{\braopket}[3]{\ensuremath{\langle#1\mid#2\mid#3\rangle}}

\newcommand{\unit}[1]{\ensuremath{\mathrm{#1}}}
\newcommand{\extderiv}{\ensuremath{\mathrm{d}}}

\newcommand{\fourquad}{\ensuremath{\quad\quad\quad\quad}}
\newcommand{\threequad}{\ensuremath{\quad\quad\quad}}

\numberwithin{equation}{section}

\title{Geometric treatment of conduction electron scattering by crystal lattice strains and dislocations}

\author[1,2]{Koushik Viswanathan\thanks{kviswana@purdue.edu} }
\author[2]{ Srinivasan Chandrasekar}
\affil[1]{Department of Physics\\Purdue University, West Lafayette, IN, USA}
\affil[2]{Center for Materials Processing and Tribology\\Purdue University, West Lafayette, IN, USA}

\date{ }

\begin{document}
\maketitle

\begin{abstract}
A theory for conduction electron scattering by inhomogeneous crystal lattice strains is developed, based on the differential geometric treatment of deformations in solids. The resulting fully covariant Schr\"odinger equation shows that the electrons can be described as moving in a non-Euclidean background space in the continuum limit of the deformed lattice. Unlike previous work, the formalism is applicable to cases involving purely elastic strains as well as discrete and continuous distributions of dislocations --- in the latter two cases it clearly demarcates the effects of the dislocation strain field and core and differentiates between elastic and plastic strain contributions respectively. The electrical resistivity due to the strain field of edge dislocations is then evaluated using perturbation theory and the Boltzmann transport equation. The resulting numerical estimate for Cu shows good agreement with experimental values, indicating that the electrical resistivity of edge dislocations is not entirely due to the core, contrary to current models. Possible application to the study of strain effects in constrained quantum systems is also discussed.
\end{abstract}

\section{Introduction}
\label{sec:introduction}

The problem of conduction electron scattering by lattice strains in metals has been studied for several decades \cite{Watts_DislSolids_1989}. The effects of homogeneous (spatially constant) elastic lattice strains on the electronic properties of solids are well established, based on symmetry principles \cite{BirPikus_SymmetryStrain_1974}. This approach cannot be used for inhomogeneous (spatially varying) strains and a theoretical description of the resulting effects on electronic conduction phenomena is not straightforward. Dislocations are sources of internal inhomogeneous strains in crystals. When considered as distributions (in a continuum sense), they contribute to plastic strains in the lattice \cite{Nabarro_CrystalDislocations_1967,Kroner_LesHouches_1981}. 

To treat purely elastic inhomogeneous strains, several models have been proposed for both metals and semiconductors ~\cite{BardeenShockley_PhysRev_1950, Zhang_PhysRevB_1994, LassenETAL_JMathPhys_2005}. The case of strains due to dislocations in metals was treated in detail by Hunter and Nabarro \cite{HunterNabarro_ProcRoySoc_1953}, who calculated the electrical resistivity due to edge and screw dislocations. They assumed that the free electrons in metals are perturbed by the deformation potential \cite{BardeenShockley_PhysRev_1950}. In this context, the contribution to electron scattering arises only from the dilatation components of the strain tensor. The resulting potential depends only on the Fermi energy and is independent of the electron--phonon coupling. It was further postulated that the electron effective mass depends on this coupling and estimates for the electrical resistivity of both edge and screw dislocations in Cu were obtained. Farvacque and Lenglart \cite{FarvacqueLenglart_PhysStatSolB_1977a, FarvacqueLenglart_PhysStatSolB_1977b} considered inhomogenous strains as perturbations of the unstrained crystal Hamiltonian, using a pseudopotential approach. Their expression for the scattering potential in semiconductors contains only the dilatation term and thus has no contribution for screw dislocations. Subsequently, Watts calculated metallic resistivity arising from dislocation strain fields using the dephasing method \cite{Watts_JPhysF_1988}. Like the earlier result \cite{HunterNabarro_ProcRoySoc_1953}, the conclusion was that the lattice strain causes insignificant electron scattering.


In contrast to metals, the effects of dislocations on the electronic properties of semiconductors have to be estimated using the complete band structure \cite{YouJohnson_SolidStatePhys_2009} and thus one must often resort to density functional calculations \cite{Kioseoglou_PhysStatSolB_2013}. The effect of a charged dislocation can be treated in a continuum sense by considering Coulomb scattering from a line of uniform charge density \cite{LookSizelove_PhysRevLett_1999}. In semiconductors and intermetallics, broken bonds in the core can give rise to quasilocalized states and various related optical effects \cite{YouJohnson_SolidStatePhys_2009, KontsevoiETAL_PhilMagLett_1998}. Conduction electron scattering by the dislocation strain field has also been presented for semiconductors \cite{JenaMishra_ApplPhysLett_2002}, exactly analogous to the deformation potential calculation in metals.

The effect of inhomogeneous elastic strains on the electronic band structure in semiconductors can be modeled \cite{Zhang_PhysRevB_1994, LassenETAL_JMathPhys_2005} as an extension of the coordinate transformation method \cite{BirPikus_SymmetryStrain_1974} for homogeneous strains. This is the most commonly used technique to estimate the shift in band degeneracies \cite{Linnik_JPhysCondMat_2012}. This method has also been applied to elastic media containing dislocations without a microscopic model \cite{Aurell_JPhysA_1999, SilvaFurtado_JPhysCondMat_2008}. 

In parallel, it was recognized that the strain field due to dislocations can have non-trivial topological effects \cite{Teichler_PhysLettA_1981, Rebane_PhysRevB_1995, BauschETAL_PhysRevLett_1998}. This is because the displacement field is multi-valued around the dislocation line --- a property which extends well beyond the core. As a consequence, topologically protected gapless modes can occur along screw dislocation lines in certain binary bulk materials causing them to behave as topological insulators \cite{RanETAL_NatPhys_2009}. Dislocations can also cause Berry phase effects \cite{SundaramNiu_PhysRevB_1999}, which have been observed in electron diffraction images \cite{BirdPreston_PhysRevLett_1988}.

In this work, an effective Schr\"odinger equation is obtained from a microscopic Hamiltonian to describe the effect of smooth inhomogeneous strain fields in crystals. The starting point is the same as that in Ref. \cite{BauschETAL_PhysRevLett_1998} but the method is put on firm mathematical ground and hence extended to apply to the case of more general plastic strains. The fundamental idea is that when considering nearest neighbour hopping in a tight-binding description, the topology of the underlying lattice must be naturally taken into account. This is done using the differential geometric formulation of deformations in solids \cite{MarsdenHughes_MathFoundElasticity, BilbyETAL_ProcRoySoc_1955, Kroner_LesHouches_1981}, which is briefly introduced in Sec.~\ref{sec:background}. An effective Schr\"odinger equation is derived in Sec.~\ref{sec:effective_Hamiltonian}, to describe conduction electrons in simple metals. The developed formalism is then applied to the case of an edge dislocation strain field and an estimate of the electrical resistivity due to a collection of parallel edge dislocations is obtained in Sec.~\ref{sec:conduction_electron_edge_dislocation}. In Sec.~\ref{sec:discussion}, numerical values of the resistivity are compared with experimental results and the assumptions involved in the calculation are discussed. The possibility of treating constrained quantum systems is also considered. Finally, some concluding remarks are presented in Sec.~\ref{sec:conclusions}.
\section{Background --- Geometric description of lattice strain}
\label{sec:background}

Inhomogeneous strains in crystals can be analyzed in the continuum sense using the methods of differential geometry. The case of purely elastic strains can be described by Riemmannian geometry \cite{MarsdenHughes_MathFoundElasticity}. The treatment of plastic strains due to dislocations in terms of non-Riemannian geometry was first outlined by Bilby \emph{et al.} \cite{BilbyETAL_ProcRoySoc_1955} and Kondo \cite{Kondo_RAAGMemoirs_1955} (for a review, see Refs. \cite{Kroner_LesHouches_1981, Nabarro_CrystalDislocations_1967,YavariGoriely_ArchRatMechAnal_2012}). Other formulations have also been presented, analogous to the gauge theory of gravity \cite{KatanaevVolovich_AnnPhys_1992, LazarAnastassiadis_PhilMag_2009, Lazar_JPhysA_2002}.

The basic idea is that in a deformed crystal, containing both elastic and plastic strains, two separate coordinate systems can be defined --- one moving internally along a lattice point and the other observing the deformation externally. The corresponding measurements of infinitesimal lengths and parallel transport depend on the nature of the deformation and so the presence of a defect is detected differently in the two cases. This notion is formalized in this section --- the notation used is that of Ref.~\cite{Nakahara_GeomTopoPhys}. 

In the continuum limit, the deformed crystal can be described by a manifold $M$ with a coordinate basis $\{\partial_\mu\}$ for vectors, which forms a set of smooth linearly independent vector fields, and the dual basis $\{\textrm{d}x^\mu\}$, a set of smooth one-forms on $M$. In the absence of defects, the coordinate basis describes purely elastic deformations, which are diffeomorphisms from $M$ to $\mathbb{R}^3$. The connection $\nabla$ (with corresponding connection coefficients $\Gamma_{\mu\nu}^\lambda$) determines how vectors and forms are parallel transported on $M$. $\Omega^1(M)$ and $\mathcal X(M)$ denote the set of all smooth one-form fields and vector fields respectively. The vector fields themselves can be viewed as maps $X \in \mathcal X(M)$, $X : \mathcal F(M) \to \mathcal F(M)$ over the space of smooth functions $\mathcal F(M)$ on $M$. Then the one-form fields are maps $W \in \Omega^1(M)$, $W : \mathcal X(M) \to \mathcal F(M)$. The exponential map $\exp(tX)|_p$ generates the flow $\sigma(t):\mathbb{R} \to M$ associated with $X$ and passing through a point $p \in M$. Being a vector field itself, $\exp_t : \mathcal F(M) \to \mathcal F(M)$ gives an approximation to a smooth function $f \in \mathcal F(M)$, for small $t \in \mathbb{R}$, in the neighborhood of $p$.

The torsion and curvature on the manifold are operators $T:\mathcal X(M) \times \mathcal X(M) \to \mathcal X(M)$ and $R: \mathcal X(M) \times \mathcal X(M) \times \mathcal X(M) \to \mathcal X(M)$, defined by their action on $X,Y,Z \in \mathcal X(M)$.
	\begin{align}
	\label{eqn:torsion_curvature_definition}
		T(X,Y) &= \nabla_X Y - \nabla_Y X - [X,Y]\\
		R(X,Y,Z) &= \nabla_X \nabla_Y Z - \nabla_Y \nabla_X Z - \nabla_{[X,Y]} Z
	\end{align}
where $[.\, ,.]$ denotes the commutator or Lie bracket.  Manifolds that have non-zero torsion and curvature are referred to as Riemann--Cartan manifolds, while the smaller class of manifolds with vanishing curvature are commonly called Weitzenb\"ock manifolds \cite{YavariGoriely_ArchRatMechAnal_2012}.

The metric tensor $g(X,Y)$ defines inner products between two vectors $X$ and $Y$. In the basis $\{\partial_\mu\}$, $g_{\mu\nu} = g(\partial_\mu,\partial_\nu)$ measures lengths infinitesimally on $M$. $g_{\mu\nu}$ is related to the infinitesimal strain tensor $\epsilon_{\mu\nu}$ as $g_{\mu\nu} = \delta_{\mu\nu} - 2\, \epsilon_{\mu\nu}$. Also, the inverse of $g_{\mu\nu}$ is denoted by $g^{\mu\nu}$.

If two vectors $X$ and $Y$ have the same inner product when parallel transported (with respect to $\nabla$), then the connection is metric compatible, i.e. $(\nabla_\kappa g)_{\mu\nu} = 0$. From this it is clear that given $g$, one can obtain unique, symmetric connection coefficients $\tilde{\Gamma}_{\mu\nu}^\lambda$ --- the Christoffel symbols. If the metric tensor is constant in a particular coordinate system $\{\partial_\mu\}$, $\tilde{\Gamma}_{\mu\nu}^\lambda$ vanish identically \cite{Nakahara_GeomTopoPhys}. 

The non-coordinate basis (or the triad field) $\{\hat{e}_i\}$ and one-forms $\{\hat{\theta}^i\}$, indexed by latin letters, can be expressed as a linear combination of $\{\partial_\mu\}$ and $\{\extderiv x^\nu\}$ ($i,k, \mu,\nu = 1,2,3$; summation implied)
	\begin{align}
	\label{eqn:triad_components}
		\hat{e}_i &= e\indices{_i^\mu} \partial_\mu \quad\quad\quad \partial_\nu = e\indices{^k_\nu} \hat{e}_k\\
		\hat{\theta}^k &= e\indices{^k_\mu} \textrm{d}x^\mu \quad\quad \textrm{d}x^\nu = e\indices{_k^\nu} \hat{\theta}^k
	\end{align}
In addition, $\{\hat{e}_i\}$ can be made orthonormal with metric components $\delta_{ij}$. In this basis the connection coefficients are zero, but $T$ and $R$, being true tensors, have non-zero components. The corresponding components of the metric $(g_{\mu\nu}, \delta_{ij})$ are related by
	\begin{equation}
	\label{eqn:metric_components_triad}
		\delta_{ij} = g_{\mu\nu} e\indices{_i^\mu} \, e\indices{_j^\nu} \quad\quad\quad g_{\mu\nu} = \delta_{ij}e\indices{^i_\mu}\, e\indices{^j_\nu}
	\end{equation}

\begin{figure}
		\centering
		\includegraphics[scale=0.6]{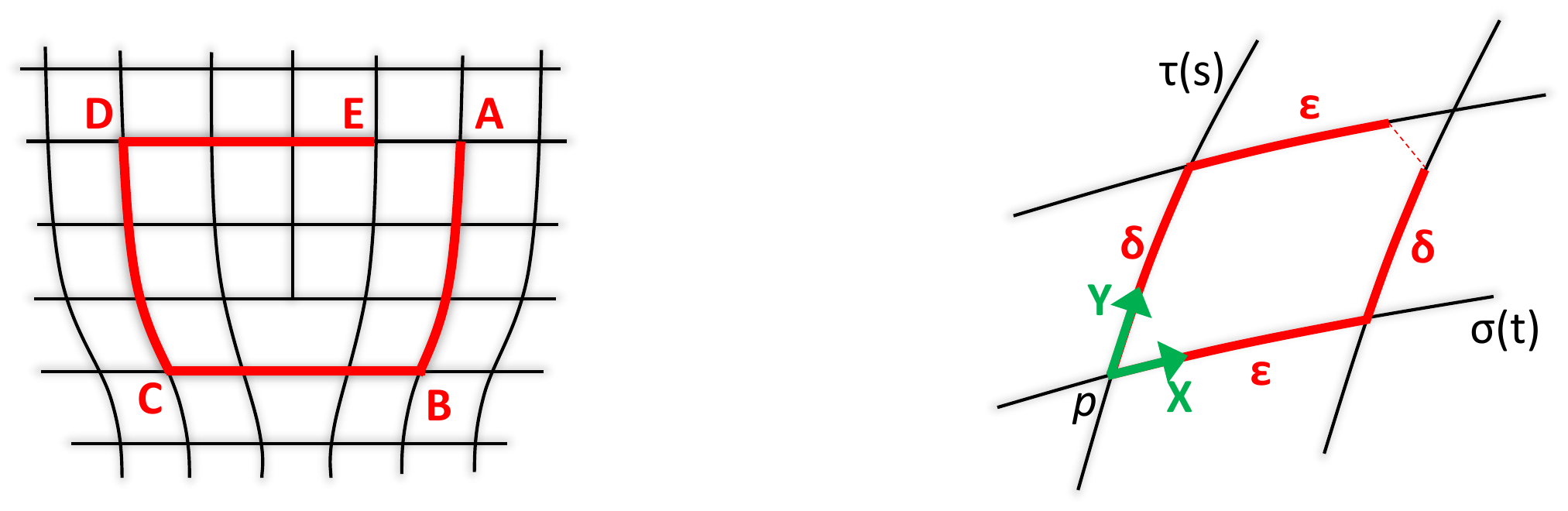}
		\caption{\label{fig:BurgersLie}Comparison between conventional Burgers' circuit and the Lie bracket: (Left) Non-closure of the Burgers' circuit $ABCDE$ indicates the presence of the edge dislocation. (Right) Non-vanishing Lie bracket $[X,Y]$ at $p$ results from an infinitesimal dislocation.}
	\end{figure}

$\{\partial_\mu\}$ and $\{\hat{e}_i\}$ (also referred to as Cartan's moving frames) correspond to the external and internal coordinates mentioned earlier. The essential difference between them is that in the presence of a dislocation the former has a vanishing Lie bracket while the latter does not. Compare the usual Burgers' circuit in a crystal containing a single edge dislocation with the definition of the Lie bracket (Fig.\ref{fig:BurgersLie}). The latter, defined as the commutator $[X,Y]$ of two vector fields $X, Y,$ measures the non-closure in travelling along the two flows $\sigma(t)$ and $\tau(s)$ generated by $X$ and $Y$ respectively, through $p$. Formally, it is the difference between traversing infinitesimal distances $\epsilon$ and $\delta$ along $\sigma(t)$ and $\tau(s)$ and in reverse order.

The components of $T$ in the $\{\partial_\mu\}$ and $\{\hat{e}_i\}$ bases are, from Eq.(\ref{eqn:torsion_curvature_definition})
	\begin{align}
	\label{eqn:torsion_global_components}
		T\indices{_{\mu\nu}^\lambda} e_\lambda &= \nabla_\mu e_\nu - \nabla_\nu \partial_\mu \equiv (\Gamma_{\mu\nu}^\lambda  - \Gamma_{\nu\mu}^\lambda) e_\lambda\\
	\label{eqn:torsion_local_components}
		T\indices{_{ij}^k} \hat{e}_k &= -[\hat{e}_i,\hat{e}_j]^k \hat{e}_k
	\end{align}
since covariant derivatives in the $\{\hat{e}_i\}$ basis commute due to zero connection coefficients. 

Eqs.(\ref{eqn:torsion_global_components}) and (\ref{eqn:torsion_local_components}) show the difference between the external and internal coordinates. The former can detect incompatible deformations (leading to torsion) on the manifold by the antisymmetry of $\Gamma^\lambda_{\mu\nu}$. The latter locally measures the non-commutativity of the flows generated by $\{\hat{e}_i\}$ via parallel transport. This is exactly analogous to the Burgers' circuit, although now it is done at every point on $M$. It is clear from this that if the non-coordinate basis has zero Lie bracket everywhere, the two coordinate systems are related by a diffeomorphism and the body has no plastic strains. 

Being antisymmetric in the two lower indices, the torsion can be thought of as a vector valued two-form $T^i$. In terms of the connection one-form $\omega \indices{_i^j} (\equiv \Gamma_{k i}^j \,\hat{\theta}^k)$, it obeys Cartan's structure equation:
	\begin{equation}
	\label{eqn:Cartan_structure_eqn}
		T^i = \textrm{d}\hat{\theta}^i + \omega\indices{_k^i}\hat{\theta}^k
	\end{equation}
where $\textrm{d}$ is the exterior derivative. $\omega\indices{_k^i}$ vanishes for the defined $\{\hat{e}_i\}$,  and the torsion tensor is an exact two-form. It can thus be integrated over a small area $D$, giving the relation
	\begin{equation}
	\label{eqn:torsion_Burgers_vector}
		b^i = \iint_D \, \sqrt{g} \, \textrm{d}x^\mu \wedge \textrm{d}x^\lambda \, T\indices{_{\mu\lambda}^i}
	\end{equation}
which defines the Burgers' vector components $b^i$ of the dislocation contained inside $D$. Thus the torsion tensor is a measure of the local dislocation density \cite{BilbyETAL_ProcRoySoc_1955, Kroner_LesHouches_1981}.

\section{Electrons in the presence of inhomogeneous strains}
\label{sec:effective_Hamiltonian}

We begin our formulation by treating the conduction electrons as independent and moving in the field of the atomic cores, which form a lattice containing $N$ unit cells. The usual Born--von K\'arm\'an periodic boundary conditions \cite{AshcroftMermin_SolidStatePhys_1980} are used, only insofar as the points in $\mathbf{k}$-space are concerned. The Hamiltonian expressed in terms of the creation and annihilation operators for Wannier states \cite{AltlandSimmons_CondMatFieldTheory} is:
	\begin{equation}
	\label{eqn:second_quantized_Hamiltonian}
		\hat{\mathcal H} = \sum_{\mathbf{R}_l, \mathbf{R}_m} \hat{\phi}^{\dag} (\mathbf{R}_l) t(\mathbf{R}_l,\mathbf{R}_m) \hat{\phi}(\mathbf{R}_m)
	\end{equation}
where the double summation is over all lattice vectors $\mathbf{R}_l, \mathbf{R}_m$. $\hat{\phi}^\dag(\mathbf{R}_l)$ ($\hat{\phi}(\mathbf{R}_m)$) creates (annihilates) a particle in the Wannier state $\ket{\mathbf{R}_l}$ ($\ket{\mathbf{R}_m}$) centered on the atom at lattice point $\mathbf{R}_l$ ($\mathbf{R}_m$). The operators obey the anti-commutation relations:
	\begin{equation}
	\label{eqn:commutation_relations}
		\{\hat{\phi}(\mathbf{R}_l),\hat{\phi}^\dag(\mathbf{R}_m)\} = \delta_{\mathbf{R}_l,\mathbf{R}_m} \threequad \{\hat{\phi}(\mathbf{R}_l),\hat{\phi}(\mathbf{R}_m)\} = \{\hat{\phi}^{\dag}(\mathbf{R}_l),\hat{\phi}^{\dag}(\mathbf{R}_m)\} = 0
	\end{equation}
$\hat{\phi}^\dag$ and $\hat{\phi}$ can be expanded in terms of the creation/annihilation operators for Bloch states $\ket{\mathbf{k}, n}$ in the perfect crystal as:
	\begin{equation}
	\label{eqn:Bloch_expansion_Wannier}
		\hat{\phi}^\dag(\mathbf{R}_m) = \frac{1}{\sqrt{N}} \sum_{\mathbf{k}, n} \braket{\mathbf{k}, n}{\mathbf{R}_m} a^\dag_{\mathbf{k},n} \fourquad \hat{\phi}(\mathbf{R}_m) = \frac{1}{\sqrt{N}} \sum_{\mathbf{k}, n} \braket{\mathbf{R}_m}{\mathbf{k}, n} a_{\mathbf{k},n}
	\end{equation}
The summations are over all relevant bands. The Hamiltonian is assumed to be diagonal in the Bloch representation, in which case $t(\mathbf{R}_l,\mathbf{R}_m)$ is given in terms of the band energy $\epsilon_n(\mathbf{k})$
	\begin{equation}
	\label{eqn:transfer_function}
		t(\mathbf{R}_l,\mathbf{R}_m) = \frac{1}{N} \sum_{\mathbf{k}, n} e^{i \mathbf{k}\cdot (\mathbf{R}_l - \mathbf{R}_m)} \epsilon_n(\mathbf{k})
	\end{equation}
If only nearest neighbour interactions are considered in Eq.(\ref{eqn:second_quantized_Hamiltonian}), then using Eq.(\ref{eqn:transfer_function}) for lattice vectors $\{\mathbf{c}_i\}$ we obtain:
	\begin{equation}
		\hat{\mathcal H} = \sum_{\mathbf{R}_l} \sum_{\mathbf{c}_i} \hat{\phi}^{\dag} (\mathbf{R}_l) \, t_{\mathbf{c}_i}(\mathbf{R}_l) \, \hat{\phi}(\mathbf{R}_l + \mathbf{c}_i)
	\end{equation}
The following simplifying assumptions are made. In view of our main application to metals, only one parabolic conduction band (with effective mass $m_e$) is considered. The band index $n$ will thus be dropped and $\sum_{\mathbf{k}}$ will range over the first Brillouin Zone (B.Z). Also, a spherical Fermi surface is assumed and a simple cubic lattice is treated. It will be seen later that these two assumptions are not critical.

For treating crystals with inhomogeneous deformation, we resort to the geometric description outlined earlier. When treated as part of a continuum, each point of the lattice $\mathbf{R}_l$ becomes a point $x$ on the body manifold $M$. The axes of the lattice can be taken to describe the coordinate basis forming a set of smooth vector fields on the manifold. Locally however, in order to move to the nearest neighbor, one has to traverse a length equal to the interatomic distance $c$ along one of the axes $\hat{e}_i$. The lattice vector $\mathbf{c}_i$ has components:
	\begin{equation}
		\mathbf{c}_i \to c\, \hat{e}_i =  c\, e\indices{_i^\mu} \partial_\mu \fourquad \mathbf{R}_l \to x \in M
	\end{equation}
In the continuum limit, $c \to 0$, expanding $t_{\mathbf{c}_i}(\mathbf{R}_l)$ in powers of $c$, 
	\begin{equation}
		t_{\mathbf{c}_i}(\mathbf{R}_l) \simeq t_0 + c \, t_1 + \frac{c^2}{2} \, t_2 + \cdots
	\end{equation}
where
	\begin{equation}
	\label{eqn:t_values}
		t_0 = \frac{1}{N} \sum_{\mathbf{k}}^{B.Z} \epsilon(\mathbf{k}) \quad\quad t_1 = \frac{ic}{N} \sum_{\mathbf{k}}^{B.Z} \epsilon(\mathbf{k}) e\indices{_i^\mu}k_\mu \quad\quad t_2 = -\frac{c^2}{2N} \sum_{\mathbf{k}}^{B.Z} \epsilon(\mathbf{k}) e\indices{_i^\mu}e\indices{_i^\nu}k_\mu k_\nu
	\end{equation}

$\hat{\phi}$ is an operator on Fock space but as far as position dependence is concerned, $\hat{\phi} \in \mathcal F(M)$. The action of the exponential map $\exp (c\,\hat{e}_i)$ gives the values of $\hat{\phi}$ along the flow generated by $\hat{e}_i$ at $x$. For small $c$, the result can be expressed in the basis $\{\partial_\mu\}$ at $x$
	\begin{equation}
	\label{eqn:operator_expansion}
		\hat{\phi}(\mathbf{R}_l + \mathbf{a}_i) \to \hat{\phi}(x) + c e\indices{_i^\mu} \partial_\mu \hat{\phi} + \frac{c^2}{2} e\indices{_i^\mu}\partial_\mu(e\indices{_i^\nu}\partial_\nu) \hat{\phi} + \cdots = \exp(c\, \hat{e}_i)|_x[\hat{\phi}]
	\end{equation}
The derivatives $\partial_\mu$ in Eq.(\ref{eqn:operator_expansion}), evaluated at $\mathbf{R}_l$, are not well defined until the limit $c \to 0$ is taken, but are formally retained. 

Linear terms in $\hat{\mathcal H}$ cancel on summation over $i$ due to inversion symmetry. Terms upto second order in $a$ are retained
	\begin{equation}
	\label{eqn:Hamiltonian_second_order}
		\hat{\mathcal H} = \sum_{\mathbf{R}_l} \hat{\phi}^\dag(\mathbf{R}_l) t_0 \hat{\phi}(\mathbf{R}_l) + \frac{c^2}{2}\sum_{\mathbf{R}_l}\sum_i \hat{\phi}^\dag(\mathbf{R}_l) t_0 e\indices{_i^\mu}\partial_\mu(e\indices{_i^\nu} \partial_\nu)\hat{\phi} - \frac{c^2}{2} \sum_{\mathbf{R}_l}\sum_i \hat{\phi}^\dag(\mathbf{R}_l) t_2 \hat{\phi}(\mathbf{R}_l)
	\end{equation}
The first term is a constant energy shift and is ommitted in the following. 

Using the assumptions stated earlier, the value of $t_0$ in Eq.(\ref{eqn:t_values}) is estimated by converting the sum in $\mathbf{k}$-space to an integral and using $V = N c^3$
	\begin{equation}
	\label{eqn:t0_estimate}
		t_0 \approx -\frac{1}{N} \left(\frac{V}{8 \pi^3}\right)\iiint_{-\pi/c}^{\pi/c} dk_x dk_y dk_z \frac{\hbar^2}{2 m_e}(k_x^2 + k_y^2 + k_z^2) = \frac{\hbar^2}{2 \tilde{m}_e c^2}
	\end{equation}
$\tilde{m}_e$ is an altered effective mass\footnote{$\tilde{m}_e$ arises because of the continuum limit of a discrete model, where the basis wavefunctions are centered on the lattice sites. It is clearly independent of the strain in the lattice.}: $\tilde{m}_e = m_e/\pi^2$. The leading order term $t_0$ is hence $O(c^{-2})$. Since the resulting term in the Hamiltonian is $O(1)$, only the $t_0$ in Eq.~\ref{eqn:Hamiltonian_second_order} is considered henceforth.

The volume element in $\{\hat{e}_i\}$ coordinates is the the unit cell volume $c^3$. This must transform into the volume 3-form in the $\{\partial_\mu\}$ basis. In order to generalize $\delta_{ij}$ in Eq.~(\ref{eqn:commutation_relations}) to the Dirac $\delta$-function, an additional factor of $\sqrt{g}$ is needed. In the continuum limit, 	
	\begin{equation}
	\label{eqn:continuum_limit}
		\sum_{\mathbf{R}_l} c^3 \to \int_M \sqrt{g} \: \textrm{d}x^1 \wedge \textrm{d}x^2 \wedge \textrm{d}x^3 \fourquad\quad \frac{\hat{\phi}(\mathbf{R}_l)}{c^{3/2}} \to \frac{\hat{\phi}(x)}{g^{1/4}} \fourquad\quad \frac{\hat{\phi}^{\dag}(\mathbf{R}_l)}{c^{3/2}} \to \frac{\hat{\phi}^{\dag}(x)}{g^{1/4}}
	\end{equation}
$\hat{\phi}^{\dag}(x)$, $\hat{\phi}(x)$ are the field creation and annihilation operators which appear on the RHS of Eq.~(\ref{eqn:operator_expansion}). Using these relations, the continuum version of the Hamiltonian is obtained by taking $\lim c\to 0$ in Eq.(\ref{eqn:Hamiltonian_second_order})
	\begin{equation}
		\hat{\mathcal H} = \int \extderiv^3 x \left(-\frac{\hbar^2 }{2\tilde{m}_e}\right)\hat{\phi}^\dag(x)\left[ \delta^{ij} e\indices{_j^\mu} \partial_\mu(e\indices{_i^\lambda}\partial_\lambda)\right] \hat{\phi}(x)
	\end{equation}
Using the relation $\Gamma_{\mu\lambda}^\nu = -\delta^{ij}e\indices{_j^\rho} \,g_{\lambda\rho} \,\partial_\mu \,e\indices{_i^\nu}$ and Eq.(\ref{eqn:metric_components_triad}), we have the relations
	\begin{equation}
		\delta^{ij} g^{\mu\rho} e\indices{_j^\mu}\,\partial_\mu(e\indices{_i^\lambda}\partial_\lambda)  = g^{\mu\nu}\,\partial_\mu\, \partial_\nu - g^{\mu\nu}\Gamma_{\mu\nu}^\lambda\,\partial_\lambda
	\end{equation}
which leads to
	\begin{equation}
		\hat{\mathcal H} = -\frac{\hbar^2}{2 \tilde{m}_e} \int \extderiv^3 x\, \hat{\phi}^\dag(x) \left(g^{\mu\nu}\nabla_\mu \nabla_\nu \right) \hat{\phi}(x)
	\end{equation}
Using the continuum equivalent of Eq.~(\ref{eqn:commutation_relations}), the action of $\hat{\mathcal H}$ on an eigenstate $\ket{\epsilon}$ yields an effective Schr\"odinger equation for the wavefunction $\psi_{\epsilon}(x)$ 	 \begin{equation}
	\label{eqn:effective_Hamiltonian}
		\hat{\mathcal H}\psi_{\epsilon}(x) = -\frac{\hbar^2}{2\tilde{m}_e} g^{\mu\nu}\nabla_\mu \nabla_\nu \psi_{\epsilon}(x)
	\end{equation}

$\mathcal{\hat{H}}$ is a a completely covariant scalar on $M$, which reflects the arbitrariness in the choice of the unit cell. The electron can hence be thought of as moving on a Riemann--Cartan manifold formed by the atomic cores displaced from the ideal lattice sites. In the case of dislocations, this also includes the effect of the torsion in the medium, thus accounting for dislocation core and strain field effects separately. Analogously, for continuous distributions of dislocations, the elastic and plastic strain contributions are accounted for by the metric and torsion parts of $\hat{\mathcal H}$ separately --- the latter being proportional to the dislocation density (see Eq.~\ref{eqn:torsion_Burgers_vector}). When the lattice has no plastic strains, Eq.~(\ref{eqn:effective_Hamiltonian}) reduces to the Schr\"odinger equation studied by Lassen \emph{et al.} \cite{LassenETAL_JMathPhys_2005}. When the strains are constant and purely elastic, it reduces to the Hamiltonian used for homogeneous strains \cite{BirPikus_SymmetryStrain_1974}, upto a change in $\tilde{m}_e$ (see earlier footnote).

Determining the exact eigenstates of $\hat{\mathcal H}$ requires the Green's function of the Laplace--Beltrami operator on an arbitrary manifold $M$, for which no general formula exists \cite{GreenMarshall_ProcRoySoc_2013}, necessitating the use of a perturbation scheme when possible.

\section{Scattering by edge dislocations}
\label{sec:conduction_electron_edge_dislocation}

\subsection{Dependence on the strain field and dislocation core}
To apply Eq.(\ref{eqn:effective_Hamiltonian}) to the case of a single edge dislocation, the following relations are used \cite{Nakahara_GeomTopoPhys}
	\begin{align}
	\label{eqn:connection_components_breakdown}
		\Gamma_{\mu\nu}^{\alpha} = \tilde{\Gamma}_{\mu\nu}^\alpha &+ \frac{1}{2} g^{\alpha\rho}\left(T_{\mu\nu\rho} + T_{\rho\mu\nu} - T_{\nu\rho\mu}\right)\\
	\label{eqn:Laplace_Beltrami_definition}
		\frac{1}{\sqrt{g}}\partial_\mu(\sqrt{g} g^{\mu\nu} \partial_\nu \Psi) &= g^{\mu\nu}\partial_\mu \partial_\nu \Psi - g^{\mu\nu}\tilde{\Gamma}_{\mu\nu}^\lambda\partial_\lambda \Psi
	\end{align}
Expanding $g^{\mu\nu}\nabla_\mu\nabla_\nu$ using Eqs.~(\ref{eqn:connection_components_breakdown}), (\ref{eqn:Laplace_Beltrami_definition}) and the definition of $\nabla$, the time--independent Schr\"odinger equation is
	\begin{equation}
	\label{eqn:Hamiltonian_split}
		-\frac{\hbar^2}{2 \tilde{m}_e} \left[\frac{1}{\sqrt{g}} \partial_\mu(\sqrt{g} \, g^{\mu\nu}\partial_\nu) - g^{\lambda\rho} T\indices{_{\mu\rho}^\mu}\partial_\lambda\right]\psi(x) = E \, \psi(x)
	\end{equation}
For a single edge dislocation along the $z$-axis in an isotropic linear elastic medium, in cylindrical coordinates $(r,\phi,z)$:
	\begin{equation}
		g_{\lambda\rho} \mathrm{d}x^\lambda \mathrm{d}x^\rho \equiv \left(1+\frac{1-2\nu}{1-\nu} \frac{b}{2\pi r} \sin\phi\right) (dr^2 + r^2d\phi^2) - \left(\frac{b\cos\phi}{\pi(1-\nu)}\right)dr d\phi + dz^2
	\end{equation}
where $\nu$ and $b$ are Poisson's ratio and Burgers' vector magnitude respectively. The direction $\phi = 0$ coincides with the Burgers' vector. Clearly, $g_{\lambda\rho} = g^{(0)}_{\lambda\rho} + h_{\lambda\rho}$, with the Euclidean metric $g^{(0)}_{\lambda\rho}$ in $(r,\phi,z)$ coordinates and strain field contribution $h_{\lambda\rho}$. In the continuum limit, $b = O(c)$, $g$ and $g^{\lambda\rho}$ can be expanded in terms of a series in $b$ retaining only the linear term:
	\begin{align}
		\label{eqn:linear_metric_tensor_determinant}
		g &= r^2\left(1+\frac{1-2\nu}{\pi(1-\nu)} \frac{b}{r}\sin\phi\right)\\
		\label{eqn:linear_inverse_metric}
		g^{\lambda\rho} &=
			\begin{pmatrix}
				1-\frac{1-2\nu}{\pi(1-\nu)} \frac{b}{r}\sin\phi & \frac{1}{2\pi(1-\nu)} \frac{b}{r^2} \cos\phi & 0\\
				\frac{1}{2\pi(1-\nu)} \frac{b}{r^2} \cos\phi & \frac{1}{r^2}\left(1-\frac{1-2\nu}{\pi(1-\nu)} \frac{b}{r}\sin\phi\right) & 0\\
				0 & 0 & 1
				\end{pmatrix}
		\end{align}
Implicit in these relations are the facts that $\sqrt{g} \simeq r\left(1+\frac{1-2\nu}{2\pi(1-\nu)} \frac{b}{r}\sin\phi\right)$ and $g^{\lambda\rho} \equiv g_{\lambda\rho}^{-1}$ upto linear order. Using this expression and expanding the terms in Eq.(\ref{eqn:Hamiltonian_split}), we arrive at the following expression for $\hat{\mathcal H}$
	\begin{equation}
	\label{eqn:final_Hamiltonian_split}
		\hat{\mathcal H} = \hat{\mathcal H}_0 + b(\hat{V}_S + \hat{V}_C)
	\end{equation}
where $\hat{\mathcal H}_0 = -\frac{\hbar^2}{2 \tilde{m}_e} \nabla^2$ and 
	\begin{align*}
		\hat{V}_S &= \frac{\hbar^2}{2\tilde{m}_e(1-\nu)} \, \frac{1}{2\pi r} \left[(1-2\nu) \sin\phi\left(\partial_{rr} + \frac{1}{r}\partial_r + \frac{1}{r^2}\partial_{\phi\phi}\right) - \frac{2}{r}\cos\phi\:\partial_{r\phi} + \frac{1}{r^2}\cos\phi\:\partial_\phi + \frac{1}{r}\sin\phi\:\partial_r\right]\\
		\hat{V}_C &= -\frac{\hbar^2}{2\,\tilde{m}_e\,r^2}\frac{1}{2\pi}\delta(r)\partial_\phi
	\end{align*}
 $\nabla^2$ is the Laplacian in $(r,\phi,z)$ coordinates. $\hat{\mathcal H}_0, \hat{V}_S$ and $\hat{V}_C$ are operators corresponding to the free space Hamiltonian, strain field and dislocation core respectively. $\hat{V}_C$ is obtained from Eq.(\ref{eqn:torsion_Burgers_vector}) by noting that a single edge dislocation with Burgers' vector $b$ along the $x$-axis contains the term $\delta(r)$ in $T^i$.


\subsection{Matrix elements and scattering probability}
	\begin{figure}
		\centering
		\includegraphics[scale=0.25]{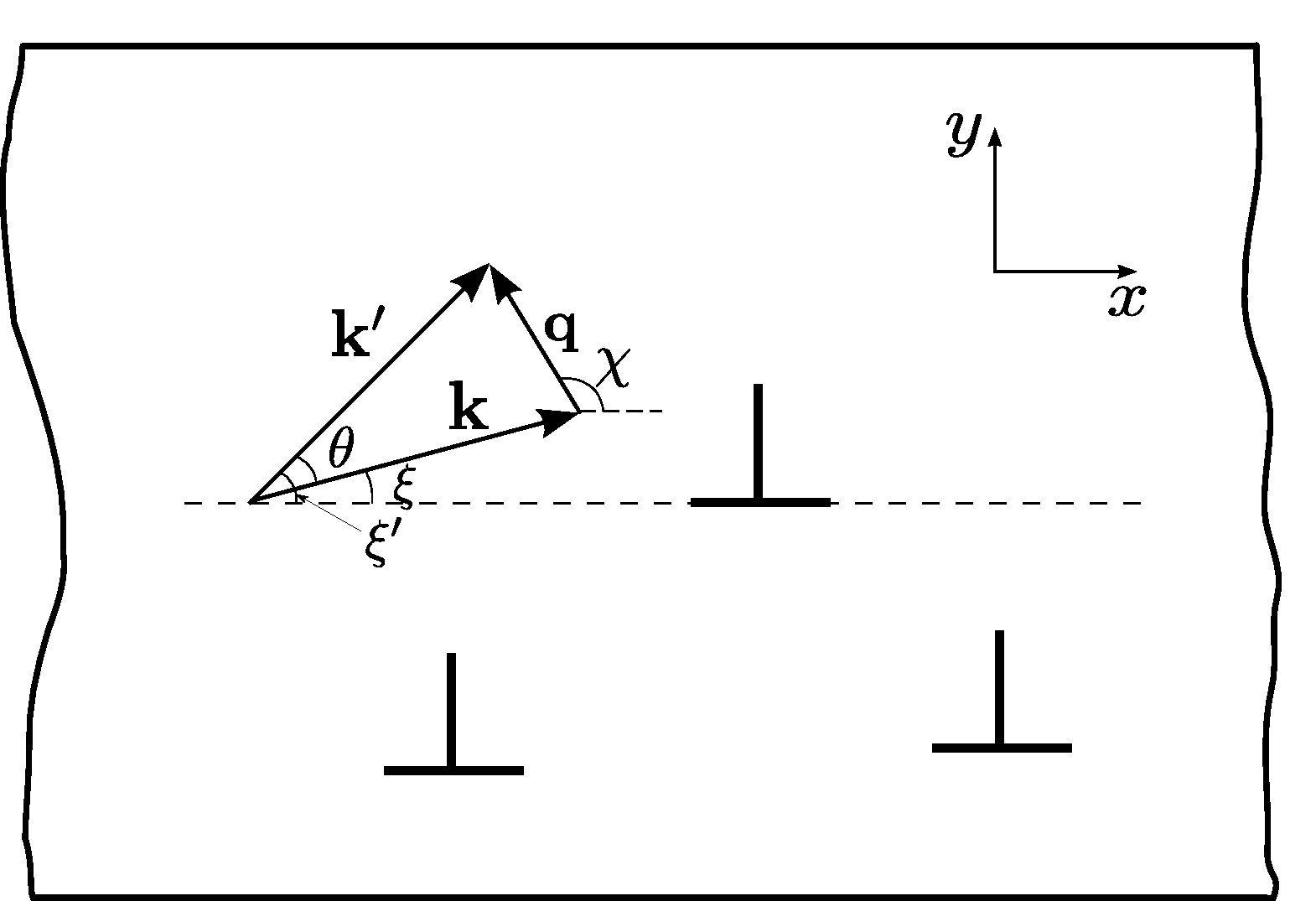}
		\caption{\label{fig:schematic}Geometry of the scattering process showing incoming and scattered wave vectors $\mathbf{k}$ and $\mathbf{k}^\prime$.}
	\end{figure}

It must be emphasized that $\hat{V}_C$ and $\hat{V}_S$ given in Eq.(\ref{eqn:final_Hamiltonian_split}) are only valid in the continuum description. The $\delta$-function in $\hat{V}_C$, which is a result of considering just a line of discontinuity in the torsion field, will become regularized when the exact atomic positions in the core are taken into account. 

Since $b$ is a small parameter, $\hat{V}_S$ is treated as a perturbation. The eigenstates of $\hat{\mathcal H}_0$, denoted by $\ket{\mathbf{k}}$ and $\ket{\mathbf{k}^\prime}$, are normalized such that:
	\begin{equation}
	\label{eqn:state_normalization}
		\braket{\mathbf{k}}{\mathbf{k}^\prime} = \delta_{\mathbf{k},\mathbf{k}^\prime} \quad\quad\quad \braket{\mathbf{r}}{\mathbf{k}} = \frac{1}{\sqrt{V}} e^{i\mathbf{k}\cdot \mathbf{r}}
	\end{equation}
where $V = L_x L_y L_z$ is a volume. A schematic of the geometry is shown in Fig.~\ref{fig:schematic}, depicting the angles $\xi,\xi^\prime$ and $\chi$. The vector $\mathbf{q} = \mathbf{k}^\prime - \mathbf{k}$ has components $q$ and $q_z$ in the plane normal to the dislocation axis and along the dislocation axis respectively. The corresponding components of $\mathbf{k}$ and $\mathbf{k}^\prime$ are $k,k^\prime$ and $k_z,k_z^\prime$ respectively. Since only elastic scattering is considered, $k = k^\prime$ and $q = 2k\sin\frac{\theta}{2}$. Also, $\chi = \frac{\pi}{2} + \xi + \frac{\theta}{2}$ and $\theta = \xi^\prime - \xi$. 

For the effect of the strain field, the corresponding matrix elements are evaluated taking an artificial core cut-off radius $a$, 
	\begin{equation}
		\braopket{\mathbf{k}^\prime}{\hat{V}_S}{\mathbf{k}}|_a = \frac{\hbar^2 }{2 \tilde{m}_e(1-\nu)} \frac{b}{2\pi}\: \frac{\sin(q_z L_z/2)}{q_z} \left[k^2 I_1 + i\, k I_2 - k^2 I_3\right]
	\end{equation}
where
	\begin{align*}
		I_1 &= -\frac{1-2\nu}{V} \int_{a}^{\infty} \int_{0}^{2\pi} dr\, d\phi \: \sin\phi\, e^{i q r\cos(\phi-\xi)} \\
		I_2 &= \frac{1}{V} \int_{a}^{\infty} \int_{0}^{2\pi} \frac{dr}{r}\: d\phi\: \sin(2\phi - \xi)\, e^{i qr \cos(\phi -\xi)} \\
		I_3 &= \frac{1}{V} \int_{a}^{\infty} \int_{0}^{2\pi} r dr\, d\phi \: \cos\phi\: \sin(2\phi - 2\xi)\, e^{i qr \cos(\phi -\xi)}
	\end{align*}
The integrals are simplified using special functions to yield
	\begin{equation}
	\label{eqn:perturbation_matrix_elements_a}
		\braopket{\mathbf{k}^\prime}{\hat{V}_S}{\mathbf{k}}|_a = -\frac{\hbar^2\, b\,k}{2\, \tilde{m}_e \, V} \frac{\sin(q_z L_z/2)}{q_z} \, i \, H(\xi,\theta,a)
	\end{equation}
with $H(\xi,\theta,a)$ given by
	\begin{equation}
		H(\xi,\theta,a) = \left[\left(\frac{1-2\nu}{1-\nu}\right) \, \frac{\cos(\theta/2 + \xi)}{\sin(\theta/2)} + \frac{\cos\theta}{\sin(\theta/2)} \frac{\cos(\xi + \theta/2)}{1-\nu}\right]J_0(q a) - \frac{2\sin(\theta + \xi)}{1-\nu} \, \frac{J_1(qa)}{qa}
	\end{equation}
$J_0(x)$ and $J_1(x)$ are Bessel functions of the first kind of order 0 and 1 respectively.

In our formulation, the effect of the strain field is taken in a continuum sense when $a \to 0$. It is only under these conditions that the field operators in Eq.(\ref{eqn:continuum_limit}) are recovered. Also, $\hat{V}_S$ can be considered a perturbation only when $b$ is very small, which implies that the integrals for the matrix elements must have lower limit $a=0$. Taking $\lim a\to 0$ the final matrix elements are:
	\begin{equation}
	\label{eqn:perturbation_matrix_elements}
		\braopket{\mathbf{k}^\prime}{\hat{V}_S}{\mathbf{k}} = -\frac{\hbar^2\, b\,k}{2\, \tilde{m}_e \, V} \frac{\sin(q_z L_z/2)}{q_z} \, i \left[\left(\frac{1-2\nu}{1-\nu}\right) \, \frac{\cos(\theta/2 + \xi)}{\sin(\theta/2)} - \frac{\sin(\theta + \xi)}{1-\nu} + \frac{\cos\theta}{\sin(\theta/2)} \frac{\cos(\xi + \theta/2)}{1-\nu}\right]
	\end{equation}

The scattering probability rate $W_{k,k^\prime}$ for the elastic scattering process is determined from Fermi's golden rule. In order to explicitly evaluate the anisotropy due to dislocation strain field scattering, a configuration of parallel edge dislocations is considered with same Burgers' vector orientation, but distributed randomly in the sample (see Fig.~\ref{fig:schematic}). If a volume $V$ contains $N_d$ dislocations and they are sufficiently far apart so that the individual scattering events are uncorrelated, then the total scattering probability rate is given by
	\begin{equation}
	\label{eqn:scattering_probability_rate}
		W_{k,k^\prime} = \frac{2\pi}{\hbar} \, N_d |\braopket{\mathbf{k}^\prime}{\hat{V}_S}{\mathbf{k}}|^2 \, \delta(\epsilon(\mathbf{k}) - \epsilon(\mathbf{k}^\prime))
	\end{equation}
At high dislocation density, it is easier to consider a continuous distribution of dislocations, starting from Eq.~(\ref{eqn:effective_Hamiltonian}).

	\begin{figure}
		\centering
		\subfloat[$|\braopket{\mathbf{k}^\prime}{\hat{V}_S}{\mathbf{k}}|^2$]{\label{fig:scatterprob_St}\includegraphics[scale=0.5]{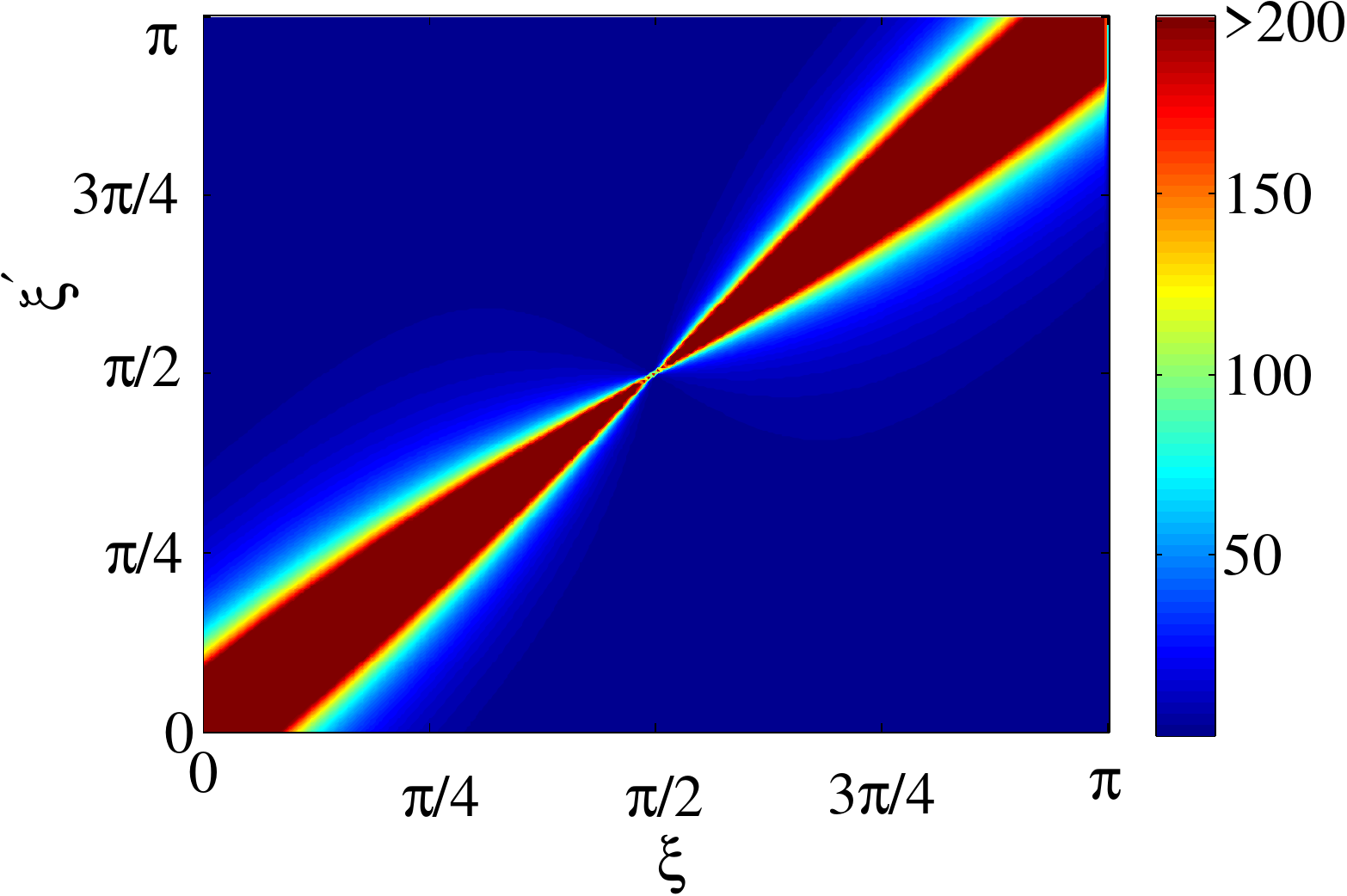}}$\quad$
		\subfloat[$|\braopket{\mathbf{k}^\prime}{\hat{U}_D}{\mathbf{k}}|^2$]{\label{fig:scatterprob_DP}\includegraphics[scale=0.5]{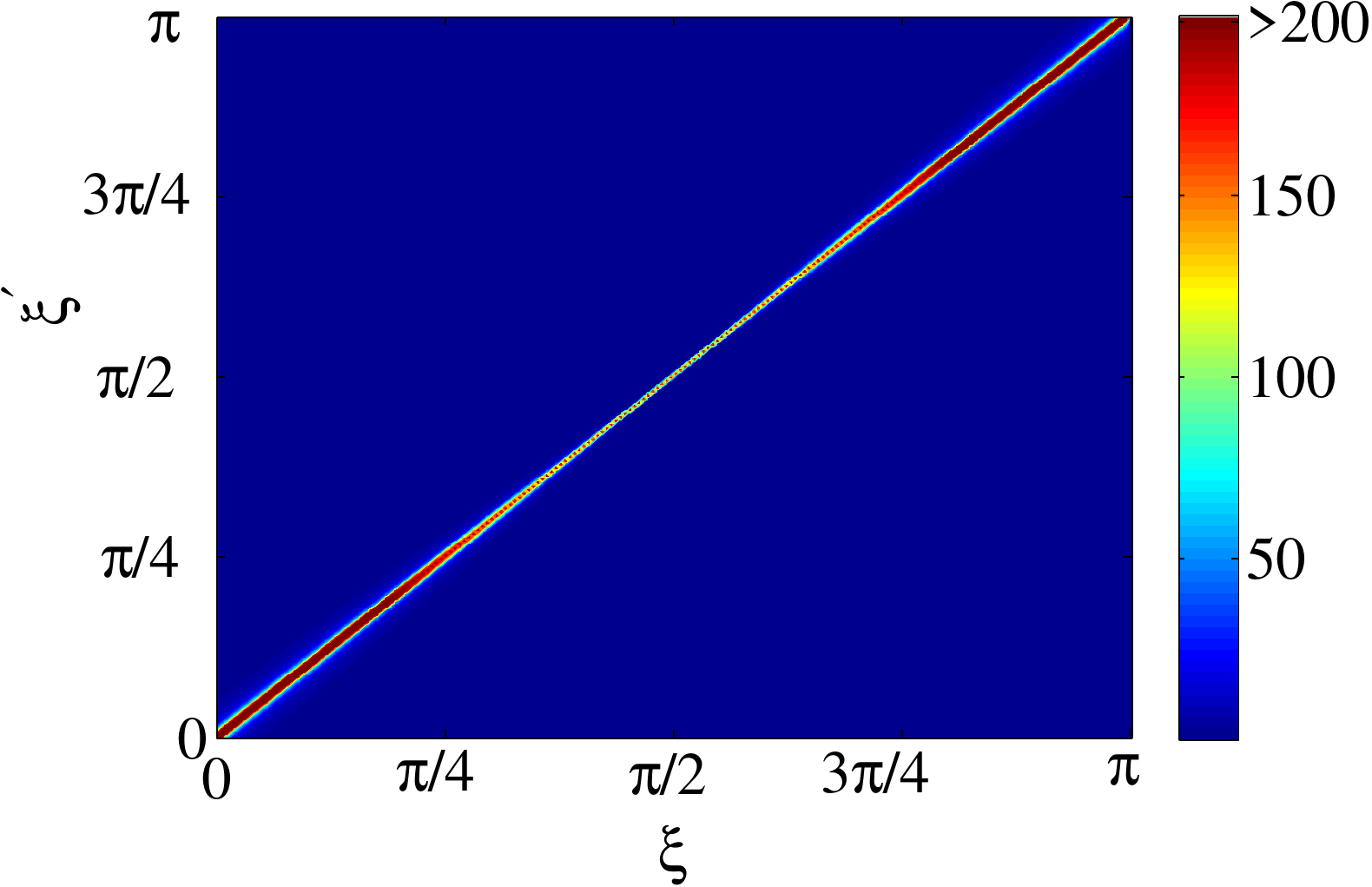}}
		\caption{Comparison of transition probability for the strain field operator $\hat{V}_S$ and deformation potential $\hat{U}_D$ as a function of incident $\xi$ and outgoing angle $\xi^\prime$ (colour bar is in arbitrary units). When compared to (b), the change from small angle scattering and enhanced anisotropy is evident in (a).}	
	\end{figure}

Other treatments of conduction electron scattering by dislocation strain-fields \cite{HunterNabarro_ProcRoySoc_1953, JenaMishra_ApplPhysLett_2002} use the deformation potential method. The change in the Fermi level at each point of the medium due to the presence of the inhomogeneous strain gives rise to an additional potential term in the Hamiltonian (See p.615 of Ref.\cite{Nabarro_CrystalDislocations_1967}) 
	\begin{equation}
	\label{eqn:deformation_potential}
		\hat{U}_D(\mathbf{r},b) = -\frac{4}{15} \frac{\hbar^2 k_F^2}{2 m_e} \Delta
	\end{equation}
where $k_F$, $\Delta$ are Fermi wave-vector and strain dilatation. 

If $\hat{U}_D$ is used instead of $\hat{V}_S$ in the same dislocation configuration to estimate the total scattering rate, fundamental differences are seen in the resulting expressions (compare Figs.~\ref{fig:scatterprob_St} and (\ref{fig:scatterprob_DP}). It is clear that the deformation potential always results in small angle scattering, as has been shown and discussed earlier \cite{HunterNabarro_ProcRoySoc_1953, BergmannETAL_PhysRevB_1981}. $\hat{V}_S$ however has a significant scattering probability away from the line $\xi = \xi^\prime$. This dependence on intermediate scattering angles \emph{a posteriori} justifies the assumption of a spherical Fermi surface in Sec.~\ref{sec:effective_Hamiltonian} --- small angle scattering results in sensitive dependence on the shape of the Fermi surface \cite{BergmannETAL_PhysRevB_1981}.

Also, the varying size of the lobe along $\xi = \xi^\prime$ indicates anisotropy as far as incident angle dependence (with respect to the glide plane) is concerned. One would thus expect the glide plane resistivity to be different from the out-of-plane value. It is precisely due to this anisotropy that the resistivity cannot be calculated using the relaxation time approximation \cite{Ziman_ElectronsPhonons_1960}. Finally, the expression for $W_{k,k^\prime}$ is symmetric with respect to $\xi$ and $\xi^\prime$, implying that $W_{k,k^\prime} = W_{k^\prime, k}$. The more general condition for this is that $\hat{V}_S$ in Eq.~(\ref{eqn:final_Hamiltonian_split}) must be Hermitian, which is shown in Appendix~\ref{app:Hermitian}.

Finally, it must be emphasized that the effect of the strain field is accompanied by that of the dislocation core too, which is treated here as a line singularity. Various analytical models for scattering from a finite dislocation core have also been considered, although in different contexts \cite{KarolikLuhvich_JPhysCondMat_1994,XuETAL_ApplPhysLett_2009}.


\subsection{Specific resistivity due to edge dislocation strain field}
The electrical resistivity due to the strain field of edge dislocations is calculated by solving the Boltzmann transport equation with an iterative technique (see Appendix \ref{app:Boltzmann}). Semiclassically, the evolution of the electron distribution function $f(\mathbf{r},\mathbf{k},t)$ in the presence of a weak uniform external electric field $\mathbf{E} = (E_x,E_y,E_z)$ is given by:
	\begin{equation}
	\label{eqn:Boltzmann_transport_equation}
		\left[\frac{df}{dt}\right]_{coll} = \frac{\partial f}{\partial t} + \frac{e \mathbf{E}}{\hbar} \cdot \nabla_k f
	\end{equation}

When $W_{k,k^\prime} = W_{k^\prime, k}$ and the system is not far from equilibrum, the function $f(\mathbf{r},\mathbf{k},t)$ can be expressed as a small deviation from the equilibrium (Fermi--Dirac) distribution $f_0$. Furthermore, we assume that the contribution from other scattering sources (phonons, impurities and vacancies) to the LHS of Eq.~(\ref{eqn:Boltzmann_transport_equation}) can be described by an isotropic relaxation time $\tau$. Under these conditions, $f$ can be written as (see Eq.~(\ref{eqn:g_corrections})):
	\begin{equation}
		\label{eqn:distribution_near_eqilibrium}
		f \simeq f_0 +\frac{e\hbar\tau}{\tilde{m}_e} \frac{\partial f_0}{\partial E} \left[\mathbf{E}\cdot \mathbf{k} + \frac{V\, \tau}{(2\pi)^3} \int d\mathbf{k}^\prime W_{k^\prime,k} \left(\mathbf{E} \cdot (\mathbf{k} - \mathbf{k}^\prime)\right) + ... \, \right]
	\end{equation}
The currents in the $x$, $y$ and $z$ directions are obtained by summing the electron velocity over all occupied states (see Eq.~(\ref{eqn:current_expressions})). Considering the leading two terms of the series in Eq.~(\ref{eqn:distribution_near_eqilibrium}) and using Eqs.~(\ref{eqn:perturbation_matrix_elements}) and (\ref{eqn:scattering_probability_rate}), the current is:
	\begin{equation}
		\label{eqn:current_simplified}
		J_x = \sigma_i E_x + \gamma_0 \left[E_x I_{xx} + E_y I_{xy}\right] \quad\quad\quad	J_y = \sigma_i E_y + \gamma_0 \left[E_x I_{yx} + E_y I_{yy}\right] \quad\quad\quad	J_z = \sigma_i E_z
	\end{equation}
with
	\begin{align}
		\label{eqn:current_exp_relations}
		\sigma_i &= \frac{n\, e^2\, \tau}{\tilde{m}_e} \quad\quad\quad\quad \gamma_0 = \frac{3}{32\pi}\frac{n \, e^2\, \tau^2}{(\tilde{m}_e)^2}\, \hbar b^2 {k_F}^2 n_d\\
		\notag
		I_{xx} &= -\int_0^{2\pi} \int_0^{2\pi} d\xi \, d\xi^\prime \Biggl[G(\xi,\xi^\prime) \sin\left(\frac{\xi+\xi^\prime}{2}\right) \sin\left(\frac{\xi^\prime - \xi}{2}\right)\cos\xi\Biggr]\\
		\notag
		I_{xy} &= \int_0^{2\pi} \int_0^{2\pi} d\xi \, d\xi^\prime \Biggl[G(\xi,\xi^\prime) \cos\left(\frac{\xi+\xi^\prime}{2}\right) \sin\left(\frac{\xi^\prime - \xi}{2}\right)\cos\xi\Biggr]\\
		\notag
		I_{yx} &= -\int_0^{2\pi} \int_0^{2\pi} d\xi \, d\xi^\prime \Biggl[G(\xi,\xi^\prime) \sin\left(\frac{\xi+\xi^\prime}{2}\right) \sin\left(\frac{\xi^\prime - \xi}{2}\right)\sin\xi\Biggr]\\
		\notag
		I_{yy} &= \int_0^{2\pi} \int_0^{2\pi} d\xi \, d\xi^\prime \Biggl[G(\xi,\xi^\prime) \cos\left(\frac{\xi+\xi^\prime}{2}\right) \sin\left(\frac{\xi^\prime - \xi}{2}\right)\sin\xi\Biggr]
	\end{align}
where $n_d$ is the dislocation density (total dislocation length per unit volume) and $G(\xi,\xi^\prime)$ is given by:
	\begin{equation}
		\label{eqn:G_fn_defn}
		G(\xi, \xi^\prime) = \Biggl(\cos\xi\cot\left(\frac{\xi^\prime - \xi}{2}\right) -\frac{1-2\nu}{2(1-\nu)}\sin\xi - \frac{\sin\xi^\prime}{1-\nu}\Biggr)^2
	\end{equation}
$\sigma_i$ denotes the (isotropic) conductivity due to phonons, point defects and vacancies. The integrals in Eq.~(\ref{eqn:current_exp_relations}), taken over the Fermi surface with radius $k_F$, are evaluated to be
	\begin{align}
		\label{eqn:integral_results}
		I_{xx} &= I^{||}(\nu) = \frac{\pi^2}{16}\left[8-\frac{16}{1-\nu} + \frac{9}{(1-\nu)^2}\right]\\
		\notag
		I_{yy} &= -I^{\perp}(\nu) =  -\frac{\pi^2}{16}\left[24-\frac{32}{1-\nu} + \frac{19}{(1-\nu)^2}\right]\\
		\notag
		I_{xy} &= I_{yx} = 0
	\end{align}

If it is assumed that Matthiessen's rule is valid and that the ratio of isotropic conductivity to dislocation conductivity is small, then the total conductivity $\sigma_{T}^{||(\perp)}$ in the glide plane and normal directions is given by
	\begin{equation}
		\sigma^{|| (\perp)}_T = \sigma_i - \frac{\sigma_i^2}{\sigma_{d}^{|| (\perp)}}
	\end{equation}
from which the relation for the specific dislocation resistivity (resistivity/dislocation density) can be obtained directly using Eq.(\ref{eqn:current_exp_relations})
	\begin{equation}
		\label{eqn:resistivity_final}
		R_d^{|| (\perp)} = \frac{\rho_d^{|| (\perp)}}{n_d} = \frac{3}{32\pi}\frac{\hbar b^2 {k_F}^2}{e^2\, n} I^{|| (\perp)}(\nu)	
	\end{equation}
Since all the terms in Eq.(\ref{eqn:resistivity_final}) are material properties or fundamental constants, the speficic resistivity (arising from the strain field alone) can be easily estimated.


\section{Numerical estimates and discussion}
\label{sec:discussion}

The expression for the specific resistivity is evaluated for edge dislocations in Cu. The expression for the effective Hamiltonian in Eq.~(\ref{eqn:effective_Hamiltonian}) was obtained for a simple cubic lattice --- by altering the limits in Eq.~(\ref{eqn:t0_estimate}), it is easily seen that only $m_e^*$ is altered for an FCC lattice. Since the final expression in Eq.~(\ref{eqn:resistivity_final}) is independent of the effective mass, this change is immaterial. Also, electrical properties of dislocations in Cu have been extensively studied experimentally, providing values to compare with.

The specific dislocation resistivity at low dislocation density was estimated from bending experiments to be $R_d = 2 (\pm 1) \times 10^{25}\, \unit{\Omega m}^3$ with $n_d = 4 \times 10^{11}\, \unit{\Omega m^{-2}}$ by Basinski and Dugdale \cite{BasinskiDugdale_PhysRevB_1985}. The configuration of edge dislocations in their setup is similar to that assumed in the calculation of Sec.\ref{sec:conduction_electron_edge_dislocation}. While these measurements were made at $4.2 \, \unit{K}$, the value is not expected to change at higher temperature \cite{Watts_DislSolids_1989}.

Kasumov \emph{et al.} \cite{KasumovETAL_SovPhysSolidState_1981} report a remarkable anisotropy in the resistivity of bent Cu samples, obtained by varying the current direction with respect to the dislocation axis. The specific dislocation resistivity along the dislocation lines (which are all parallel in the bending experiment) was found to be least when compared to the resistivity in the other directions. In contrast, the second set of experiments by Basinski and Dugdale \cite{BasinskiDugdale_PhysRevB_1985} (at high dislocation density) show a lack of anisotropy in crystals deformed by tension. The results are still not clear --- the lack of accounting for inhomogeneous dislocation distribution in bending in the work of Kasumov \emph{et al.} \cite{KasumovETAL_SovPhysSolidState_1981} could bias their results, while high dislocation density and dislocation substructure development in Basinski and Dugdale's tension experiments could even out any anisotropy. Also, the fact that in our calculation the edge dislocations are assumed to have perfectly parallel Burgers vectors, particularly enhances the anisotropy and experimental validation of this aspect of the calculation is a little uncertain.

In the derivation of Eq.(\ref{eqn:resistivity_final}), Matthiessen's rule is assumed to be valid. Further, the assumption that each of the dislocations act as independent scatterers results in $R_d$ being independent of $n_d$. This can be expected at the given dislocation densities. At much lower dislocation densities, the fact that $R_d$ depends on $n_d$ and $\sigma_i$ can be seen as the deviation from Matthiessen's rule \cite{PitsiFonteyn_SolidStateComm_1989}.

Hunter and Nabarro \cite{HunterNabarro_ProcRoySoc_1953} theoretically estimated the specific dislocation resistivity in Cu to be $R_d \simeq 0.07 \times 10^{-25} \unit{\Omega m^3}$, which is much lesser than the experimental value mentioned above. They also found an anisotropy of 3:1 for the resistivity. Another feature of their calculation is that a core radius is unnecessary for convergence of the integrals, just as in the calculation in this paper. 


\begin{table}
	\centering
	\caption{\label{tbl:constant_values}Values for the various parameters for Cu}
	\begin{tabular}{c|c|c|c|c}
			{Constant} & {$\nu$} & {$n$} & {$k_F$} & {$b$}\\\hline{} & {} & {} & {} & {}\\[-1.5ex]
			{Value} & 0.33 & {$8.47 \times 10^{28}\unit{m^{-3}}$} & {$1.36 \times 10^{10} \unit{m^{-1}}$} & {$3.61 \times 10^{-10} \unit{m}$}\\
	\end{tabular}
\end{table}

Using the values shown in Table~\ref{tbl:constant_values}, taken from elementary free-electron theory \cite{AshcroftMermin_SolidStatePhys_1980}, the electrical resistivity due to the strain field in the glide plane and normal to it is found to be:
	\begin{align}
	\label{eqn:resistivity_numerical_values}
		R_d^{||} &= 0.91 \times 10^{-25} \unit{\Omega m^3}\\
		R_d^{\perp} &= 4.03 \times 10^{-25} \unit{\Omega m^3}
	\end{align}
which is of the same order of magnitude as the experimental value of Basinski and Dugdale \cite{BasinskiDugdale_PhysRevB_1985}. It is also almost two orders of magnitude larger than all the values predicted earlier \cite{Watts_DislSolids_1989,HunterNabarro_ProcRoySoc_1953, Watts_JPhysF_1988}. 
	\begin{figure}
		\centering
		\subfloat[In-plane incidence]{\label{fig:inplane_incidence}\includegraphics[scale=0.4]{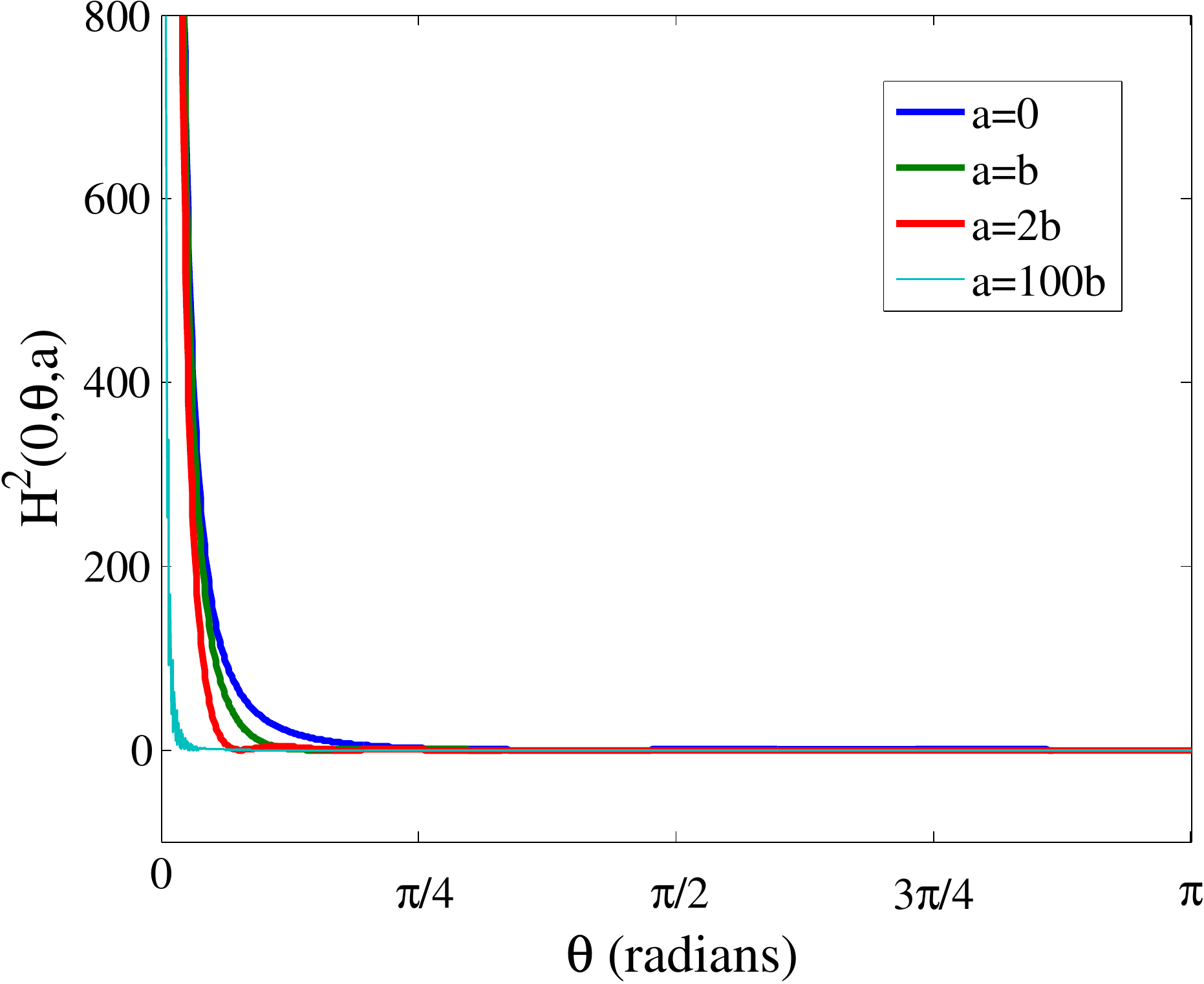}}$\quad$
		\subfloat[Normal incidence]{\label{fig:normal_incidence}\includegraphics[scale=0.4]{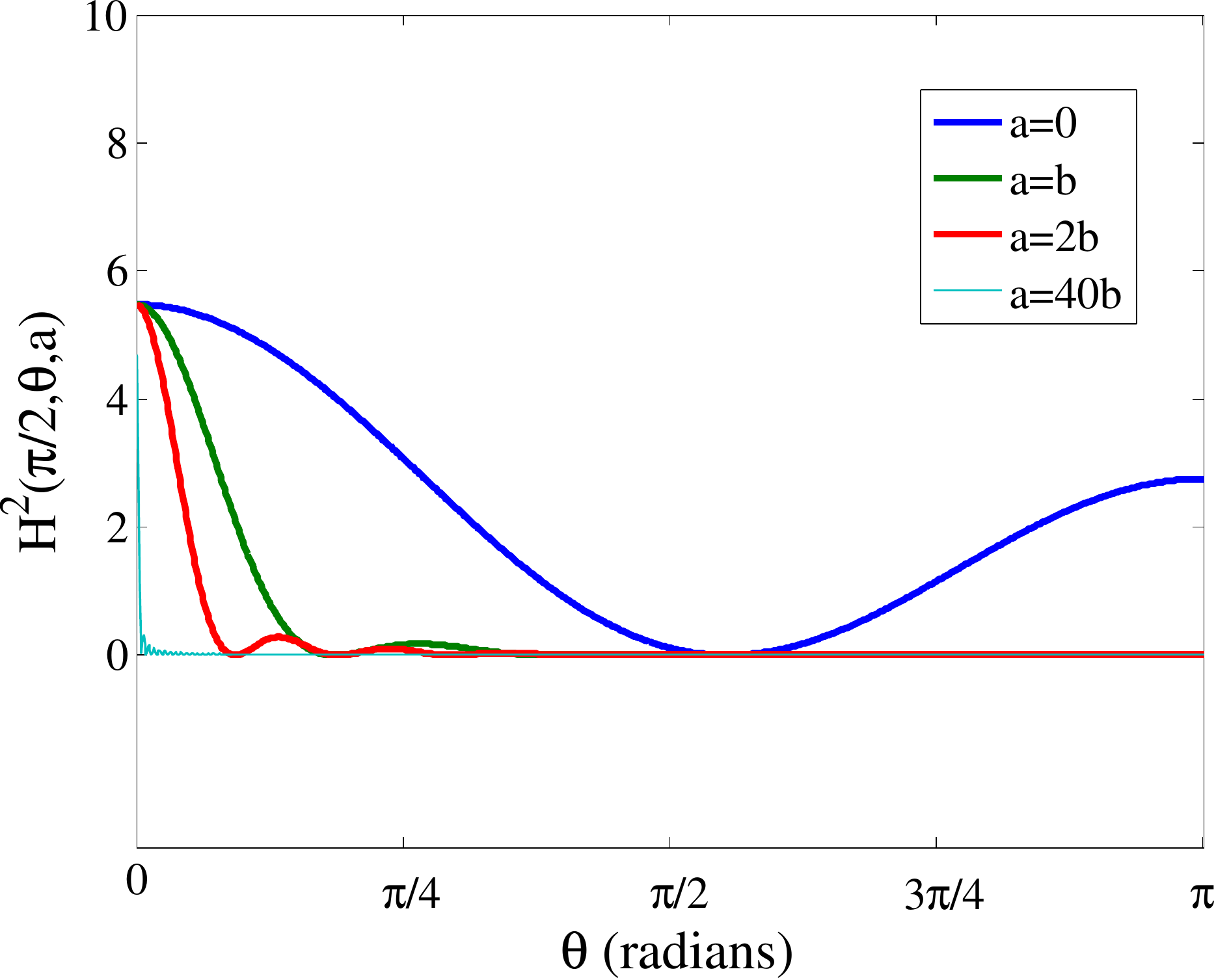}}
		\caption{Plot of variation of $(H(\xi,\theta,a))^2$ for in-plane and normal electron incidence. $y$-axis is in arbitrary units, $x$-axis is in radians.}
	\end{figure}

Calculations based on Watts' dephasing method \cite{Watts_JPhysF_1988} show that the resistivity arises from the atoms close to the dislocation line. For an artificial core cut-off radius $a = O(b)$ electron scattering is drastically reduced. The change in $H^2(\xi,\theta,a)$ (see Eq.~(\ref{eqn:perturbation_matrix_elements_a})), which is proportional to the transition probability rate, as a function of $\theta$ for varying $a$ is plotted in Fig.\ref{fig:inplane_incidence} and \ref{fig:normal_incidence} for in-plane and normal incidence ($\xi = 0$ and $\xi = \pi/2$) respectively. Finite $a$ changes the basic nature of the scattering process under normal incidence, by supressing the finite value for $\theta = \pi$. For in-plane incidence, the singularity at $\theta = 0$ is unchanged. Thus, even for incidence angles close to (but not equal to) $\pi/2$, the small-angle scattering probability is not altered significantly and one would not expect the resulting resistivity to change by much. As argued in Sec.\ref{sec:conduction_electron_edge_dislocation}, since our method is based on directly obtaining the effect of the strain field by treating the core as a line singularity, such a construct is quite misleading and the limit $a \to 0$ must be enforced. The possibility that the anisotropy in dislocation resistivity is altered for finite $a$ is shadowed by the fact that there is a non-zero contribution from the core itself, as alluded to in Sec.\ref{sec:conduction_electron_edge_dislocation}. 

The formulation in Sec.\ref{sec:effective_Hamiltonian} applies to an arbitrary strain field, independent of the presence of lattice defects. It shows that, in the continuum limit, strained lattices can be treated as a non-Euclidean medium in which the conduction electrons propagate. We now make a few remarks regarding related work. Firstly, Bausch \emph{et al.} \cite{BauschETAL_PhysRevLett_1998}, also starting from a Wannier function representation, obtained a different expression for the resulting operators (See Eq.(8) and subsequent discussion in Ref.\cite{BauschETAL_PhysRevLett_1998}). This is because of the assumption that $\tilde{\Gamma}^\lambda_{\mu\nu}$ is the symmetric part of the general connection $\Gamma^\lambda_{\mu\nu}$, which is incorrect (see Eq.~(\ref{eqn:connection_components_breakdown})). Using the definition of the non-coordinate basis it is clear that there is no need for additional partial integrations in the above formulation. The resulting term is proportional to the torsion tensor components unlike the final expression in their result, which vanishes for uniform plastic strain. Secondly, electron scattering in graphene sheets by strains \cite{Linnik_JPhysCondMat_2012} and impurities and corrugations \cite{KatsnelsonGeim_PhilTransRoySoc_2008} has been studied extensively recently. The specific problem of dislocations in graphene was studied using a covariant formalism by de Juan \emph{et al.} \cite{deJuanETAL_NuclPhysB_2010}. They commented that in bulk electronic systems, the electron wavefunction must feel the effect of torsion in the space (due to dislocations) via a coupling in the Lagrangian. Our result shows that the torsion naturally arises in the effective Schr\"odinger equation and a Lagrangian description is hence unnecessary for treating conduction electrons in metals.  

It must also be remembered that the notion of spatial curvature in these problems arises from some appropriate continuum limit of the crystal lattice. For perfect crystals, the use of the idea of an electron gas is justified within such a continuum model (referred to as the \lq jellium model\rq\ \cite{AltlandSimmons_CondMatFieldTheory}). Hence, in the presence of inhomogeneous strains, generalizing this method by use of the covariant derivative is not \emph{a priori} justified \cite{Aurell_JPhysA_1999, LassenETAL_JMathPhys_2005, SilvaFurtado_JPhysCondMat_2008}.  Indeed, as our microscopic derivation shows, the assumptions involved are obscured by such a method.


The formulation in this paper also looks promising for treating problems of constrained quantum systems, such as thin films, where electrons are constrained to move in two dimensions on a curved surface. In this context, it has been shown \cite{FerrariCuoghi_PhysRevLett_2008} that a quantum particle constrained on a surface is governed by a Schr\"odinger equation written in curvilinear coordinates, while explicitly including the effects of any applied electric and magnetic fields. The effect of an out-of-plane strain on curved nanostructures can be modeled by writing the Schr\"odinger equation in an arbitrary coordinate system, parameterizing the constraining surface, and taking a confining potential in the orthogonal direction \cite{OrtixETAL_PhysRevB_2011}. In these systems, instead of using a confining procedure \cite{daCosta_PhysRevA_1981}, our method could allow one to start from a microscopic Hamiltonian and arrive at the effective medium description by using the method of moving frames, for a given unit cell. As the scheme is not dependent on the nature of the strains, it need not be limited to plane strain conditions. 

The electrical resistivity calculation shows that the model used for conduction electrons is sufficient to obtain an estimate that matches well with experimental values. The question of anisotropy remains unanswered though, mainly because of experimental considerations \cite{Watts_DislSolids_1989, BasinskiDugdale_PhysRevB_1985} --- obtaining randomly distributed but parallely oriented dislocation structures at low dislocation density is a difficult undertaking. Finally, there could also be (perhaps comparable) contribution to the resistivity from the dislocation core. However, as shown in this work, the effect of the strain field on scattering electrons is not entirely negligible.

\section{Conclusions}
\label{sec:conclusions}

An effective Schr\"odinger equation has been obtained in the continuum limit for the conduction electrons in parabolic conduction bands in the presence of an inhomogeneous lattice strain field. A perturbative calculation is presented for the strain field of an edge dislocation in Cu and this result is used to estimate the (anisotropic) electrical resistivity of uniformly distributed parallel edge dislocations. The resulting value agrees well with experimentally established estimates for the specific dislocation resistivity which indicates that the strain field contribution to electron scattering is not insignificant. The covariant formulation presented is applicable to the case of continuous distributions of dislocations and also looks promising for describing constrained quantum systems as well.

\section{Acknowledgements}
The authors would like to acknowledge useful and stimulating discussions with Prof. G.F.Giuliani. K.V. would like to acknowledge financial support via the Frederick N. Andrews fellowship at Purdue University.

\section*{Appendix}
\appendix
\section{Hermitian nature of $\hat{V}_S$}
\label{app:Hermitian}

In general, the scattering probability rate, as given by Fermi's Golden Rule (\ref{eqn:scattering_probability_rate}) obeys the relation $W_{k,k^\prime} = W_{k^\prime,k}$ if the corresponding operator $\hat{V}$ is such that:
\begin{equation}
	\braopket{\mathbf{k}^\prime}{\hat{V}}{\mathbf{k}} = (\braopket{\mathbf{k}}{\hat{V}}{\mathbf{k}^\prime})^*
\end{equation}
i.e. is Hermitian. This can be shown for $\hat{V}_S$ derived in Sec.~\ref{sec:conduction_electron_edge_dislocation} by deriving the adjoint operator. From Eq.~(\ref{eqn:final_Hamiltonian_split})
\begin{equation}
	\hat{V}_S(b,r,\phi)= \frac{\hbar^2 b}{4\pi m(1-\nu)}\left(\hat{I}_0 - \hat{I}_1 + \hat{I}_2 + \hat{I}_3\right)
\end{equation}
where $I_0$ is proportional to the 2D Laplacian and is clearly Hermitian. We thus only need to consider $\hat{I}_1 - \hat{I}_2 - \hat{I}_3$ given by:
\begin{equation}
	\hat{I}_1 = \frac{2 \cos\phi}{r^2}  \partial_{r \phi}\threequad \hat{I}_2 = \frac{\cos\phi}{r^3} \partial_\phi \threequad	\hat{I}_3 = \frac{\sin\phi}{r^2}  \partial_r
\end{equation}

Let $k$ and $l$ denote the set of all quantum numbers characterizing the initial and final states respectively, $\psi_k$ and $\psi_l$ be the corresponding wave functions (not necessarily plane waves):
\begin{equation*}
	\braopket{l}{\hat{I}_2}{k} = \iint \, \frac{\psi_l^*}{r^2}\cos\phi  \, \frac{\partial \psi_k}{\partial \phi}  \, dr \, d\phi \fourquad	\braopket{l}{\hat{I}_3}{k} = \iint \, \frac{\psi_l^*}{r}\sin\phi  \, \frac{\partial \psi_k}{\partial r}  \, dr \, d\phi
\end{equation*}
which, after suitable partial integration, respectively become:
\begin{equation*}
	\braopket{l}{\hat{I}_2}{k} = \iint \, \frac{\psi_l^* \psi_k}{r^2}\sin\phi  \, \, dr \, d\phi - (\braopket{k}{\hat{I}_2}{l})^* \threequad \braopket{l}{\hat{I}_3}{k} = \iint \, \frac{\psi_l^* \psi_k}{r^2}\sin\phi  \, \, dr \, d\phi - (\braopket{k}{\hat{I}_3}{l})^*
\end{equation*}
Similarly, 
\begin{equation*}
	\braopket{l}{\hat{I}_1}{k} = 2 \iint \, \frac{\psi_l^* \psi_k}{r^2}\sin\phi  \, \, dr \, d\phi + (\braopket{k}{\hat{I}_1}{l})^* - 2 (\braopket{k}{\hat{I}_2}{l})^* - 2 (\braopket{k}{\hat{I}_3}{l})^*
\end{equation*}
Adding the terms together and rearranging, we get $\braopket{l}{\hat{V}_S}{k} = \left(\braopket{k}{\hat{V}_S}{l}\right)^*$.

\section{Series solution to Boltzmann transport equation}
\label{app:Boltzmann}

As is evident from Eq.(\ref{eqn:perturbation_matrix_elements}), the scattering probability is anisotropic and a function of the incident angle. Thus for electrons scattered by edge dislocations, the collision integral cannot be approximated by a meaningful relaxation time \cite{Ziman_ElectronsPhonons_1960}. For this reason, the collision integral is expanded in powers of the relaxation time for other scattering processes (phonons, impurities and vacancies) and solved to first order. This iterative solution to the Boltzmann equation was first presented by MacKenzie and Sondheimer \cite{MackenzieSondheimer_PhysRev_1950}.

The rate at which the (non-equillibrium) distribution function changes locally is given by the Boltzmann transport equation:
	\begin{equation}
		\label{eqn:boltzmann_transport_equation}
		\left[\frac{df}{dt}\right]_{coll} = \frac{\partial f}{\partial t} + \mathbf{v} \cdot \nabla f + \frac{\mathbf{F}}{\hbar} \cdot \nabla_k f
	\end{equation}
with the collision integral:
	\begin{equation}
		\label{eqn:collision_integral}
		\left[\frac{df}{dt}\right]_{coll} =  \frac{V}{(2\pi)^3} \int d\mathbf{k}^\prime \Biggl[ W_{k^\prime,k} f(\mathbf{k}^\prime)(1-f(\mathbf{k})) -  W_{k,k^\prime} (1-f(\mathbf{k}^\prime))f(\mathbf{k})\Biggr]
	\end{equation}
This integro-differential equation becomes analytically tractable under the following conditions:
	\begin{enumerate}
		\item \emph{Reversibility condition}: If $W_{k^\prime,k} = W_{k,k^\prime}$ then $\left[\frac{df}{dt}\right]_{coll} = - \frac{V}{(2\pi)^3} \int d\mathbf{k}^\prime W_{k^\prime,k} \left[f(\mathbf{k}) - f(\mathbf{k}^\prime)\right]$
		\item \emph{System not far from equilibrium}: When the system is not far from equilibrium and the scattering processes are elastic, we can express the distribution function as $f = f_0 - \frac{\hbar}{m} \mathbf{g}(\mathbf{k})\cdot \mathbf{k} \frac{\partial f_0}{\partial E}$, where $f_0(\mathbf{k})$ is the equilibrium (Fermi--Dirac) distribution function.
		\item \emph{Isotropic background scattering}: If we further assume that all other background scattering processes can be described by an isotropic relaxation time $\tau$, the collision term contrbution from the dislocation alone can be separated
			\begin{align}
				\label{eqn:collision_term_split}
				\left[\frac{\partial f}{\partial t}\right]_{coll.} &=  -\frac{f - f_0}{\tau} + \left[\frac{\partial f}{\partial t}\right]_{disl.}\\
				\notag
				\left[\frac{\partial f}{\partial t}\right]_{disl.} &= \frac{V}{(2\pi)^3} \frac{\partial f_0}{\partial E} \int d\mathbf{k}^\prime W_{k^\prime,k} \left[\mathbf{g}(\mathbf{k})\cdot \mathbf{k} -  \mathbf{g}(\mathbf{k}^\prime) \cdot \mathbf{k}^\prime\right]
			\end{align}
	\end{enumerate}
In the absence of temperature gradients, and on application of a weak electric field $\mathbf{E}$, Eq.~(\ref{eqn:boltzmann_transport_equation}) reduces to:
	\begin{equation}
		\label{eqn:boltzmann_equation_no_T}
		-\frac{e\hbar}{m}\mathbf{E}\cdot\mathbf{k}\frac{\partial f_0}{\partial E} = \left[\frac{\partial f}{\partial t}\right]_{coll}
	\end{equation}
The current, which is proportional to the integral of the wave vector over all electron states in $\mathbf{k}$ space, can be expressed as a surface integral over the Fermi surface if we assume that only electrons on the Fermi surface are responsible for conduction phenomena in metals. The relevant expressions are:
	\begin{equation}
		\label{eqn:current_expressions}
		[J_x, J_y, J_z]^T = - \frac{3\, e\, n}{4\pi m k_F} \iint \mathbf{g}(\mathbf{k})\cdot \mathbf{k} \, \sin \theta \:\left[\sin\theta\cos\phi\, ,\, \sin\theta \sin\phi\, , \, \cos\theta\right]^T \, d\theta \, d\phi \\
	\end{equation}

$e$ and $m$ are the electron charge and mass, $k_F$ and $n$ represent the Fermi wave vector and free-electron density of the metal. A relation between the current and the applied field is obtained by evaluating $\mathbf{g}(\mathbf{k})$. Combining Eqs.~(\ref{eqn:collision_term_split}) and (\ref{eqn:boltzmann_equation_no_T}), we obtain the integral equation that $\mathbf{g}(\mathbf{k})$ satisfies:
	\begin{equation}
		\label{eqn:boltzmann_transport_final}
		-e \mathbf{E}\cdot \mathbf{k} = \frac{1}{\tau} \mathbf{g}(\mathbf{k})\cdot \mathbf{k} + \frac{V}{(2\pi)^3} \int d\mathbf{k}^\prime W_{k^\prime,k} \left[\mathbf{g}(\mathbf{k})\cdot \mathbf{k} -  \mathbf{g}(\mathbf{k}^\prime) \cdot \mathbf{k}^\prime\right]
	\end{equation}
An iterative solution to Eq.(\ref{eqn:boltzmann_transport_final}) is obtained by first assuming a series expansion for $\mathbf{g} = \mathbf{g}_0 + \mathbf{g}_1 + ...$ with increasing powers of $\tau$, then inserting this expansion into the integral equation and solving orderwise. This can be done if the second term on the RHS of Eq.~(\ref{eqn:boltzmann_transport_final}) is small. The first and second order corrections are:
	\begin{equation}
		\label{eqn:g_corrections}
		\mathbf{g}_0 \cdot \mathbf{k} = -e \tau \mathbf{E} \cdot \mathbf{k} \fourquad\quad \mathbf{g}_1 \cdot \mathbf{k} = -e\tau^2 \frac{V}{(2\pi)^3} \int d\mathbf{k}^\prime W_{k^\prime,k} \left[\mathbf{E} \cdot (\mathbf{k} - \mathbf{k}^\prime)\right]
	\end{equation}
Once $W_{k^\prime, k}$ is known, the conductivity tensor is determined to second order by Eqs.~(\ref{eqn:current_expressions}) and (\ref{eqn:g_corrections}).

\bibliography{main}
\bibliographystyle{unsrt}
\end{document}